\documentclass[%
 reprint,
superscriptaddress,
 amsmath,amssymb,
 aps,
 prl,
]{revtex4-2}

\usepackage{graphicx}
\usepackage{todonotes}
\usepackage{float}
\usepackage{mhchem}
\usepackage{dcolumn}
\usepackage{bm}
\usepackage{hyperref}
\hypersetup{
   breaklinks=true,   
   colorlinks=true,   
}

\begin{document}
\preprint{APS/123-QED}
\title{High-precision mass measurement of \textsuperscript{103}Sn restores smoothness of the mass surface}%

\author{C.~M.~Ireland}%
\email{ireland@frib.msu.edu}
\affiliation{Facility for Rare Isotope Beams, East Lansing, Michigan, 48824, USA}
\affiliation{Department of Physics and Astronomy, Michigan State University, East Lansing, Michigan 48824, USA}
\author{F.~M.~Maier}
\affiliation{Facility for Rare Isotope Beams, East Lansing, Michigan, 48824, USA}
\author{G.~Bollen}
\affiliation{Facility for Rare Isotope Beams, East Lansing, Michigan, 48824, USA}
\affiliation{Department of Physics and Astronomy, Michigan State University, East Lansing, Michigan 48824, USA}
\author{S.~E.~Campbell}%
\affiliation{Facility for Rare Isotope Beams, East Lansing, Michigan, 48824, USA}
\affiliation{Department of Physics and Astronomy, Michigan State University, East Lansing, Michigan 48824, USA}
\author{X.~Chen}
\affiliation{Facility for Rare Isotope Beams, East Lansing, Michigan, 48824, USA}
\author{H.~Erington}%
\affiliation{Facility for Rare Isotope Beams, East Lansing, Michigan, 48824, USA}
\affiliation{Department of Physics and Astronomy, Michigan State University, East Lansing, Michigan 48824, USA}
\author{N.~D.~Gamage}%
\affiliation{Facility for Rare Isotope Beams, East Lansing, Michigan, 48824, USA}
\author{M.~J.~Guti\'errez}
\affiliation{Institut f\"ur Physik, Universit\"at Greifswald, 17487 Greifswald, Germany}
\affiliation{GSI Helmholtzzentrum für Schwerionenforschung, D-64291 Darmstadt, Germany}
\author{C.~Izzo}%
\affiliation{Facility for Rare Isotope Beams, East Lansing, Michigan, 48824, USA}
\author{E.~Leistenschneider}
\affiliation{Nuclear Science Division, Lawrence Berkeley National Laboratory, Berkeley, California 94720, USA}
\author{E.~M.~Lykiardopoulou}
\affiliation{Nuclear Science Division, Lawrence Berkeley National Laboratory, Berkeley, California 94720, USA}
\author{R.~Orford}
\affiliation{Nuclear Science Division, Lawrence Berkeley National Laboratory, Berkeley, California 94720, USA}
\author{W.~S.~Porter}
\affiliation{Department of Physics and Astronomy, University of Notre Dame, Notre Dame, IN, USA}
\author{D.~Puentes}%
\affiliation{Facility for Rare Isotope Beams, East Lansing, Michigan, 48824, USA}
\affiliation{Department of Physics and Astronomy, Michigan State University, East Lansing, Michigan 48824, USA}
\author{M.~Redshaw}
\affiliation{Department of Physics, Central Michigan University, Mount Pleasant, Michigan 48859, USA}
\author{R.~Ringle}
\affiliation{Facility for Rare Isotope Beams, East Lansing, Michigan, 48824, USA}
\affiliation{Department of Physics and Astronomy, Michigan State University, East Lansing, Michigan 48824, USA}
\author{S.~Rogers}
\affiliation{Facility for Rare Isotope Beams, East Lansing, Michigan, 48824, USA}
\author{S. Schwarz}
\affiliation{Facility for Rare Isotope Beams, East Lansing, Michigan, 48824, USA}
\author{L.~Stackable}
\affiliation{Facility for Rare Isotope Beams, East Lansing, Michigan, 48824, USA}
\author{C.~S.~Sumithrarachchi}
\affiliation{Facility for Rare Isotope Beams, East Lansing, Michigan, 48824, USA}
\author{A.~A.~Valverde}
\affiliation{Physics Divison, Argonne National Laboratory, Argonne, IL, USA}
\author{A.~C.~C.~Villari}
\affiliation{Facility for Rare Isotope Beams, East Lansing, Michigan, 48824, USA}
\author{I.~T.~Yandow}%
\affiliation{Facility for Rare Isotope Beams, East Lansing, Michigan, 48824, USA}
\affiliation{Department of Physics and Astronomy, Michigan State University, East Lansing, Michigan 48824, USA}
\date{\today}%

\begin{abstract}
\noindent
As a step towards the ultimate goal of a high-precision mass measurement of doubly-magic $^{100}$Sn, the mass of $^{103}$Sn was measured at the Low Energy Beam and Ion Trap (LEBIT) located at the Facility for Rare Isotope Beams (FRIB). Utilizing the time-of-flight ion cyclotron resonance (ToF-ICR) technique, a mass uncertainty of 3.7~keV was achieved, an improvement by more than an order of magnitude compared to a recent measurement performed in 2023 at the Cooler Storage Ring (CSRe) in Lanzhou. Although the LEBIT and CSRe mass measurements of $^{103}$Sn are in agreement, they diverge from the experimental mass value reported in the 2016 version of the Atomic Mass Evaluation (AME2016), which was derived from the measured $Q_{\beta^+}$ value and the mass of $^{103}$In. In AME2020, this indirectly measured $^{103}$Sn mass was classified as a `seriously irregular mass' and replaced with an extrapolated value, which aligns with the most recent measured values from CSRe and LEBIT. As such, the smoothness of the mass surface is confidently reestablished for $^{103}$Sn. Furthermore, LEBIT's mass measurement of $^{103}$Sn enabled a significant reduction in the mass uncertainties of five parent isotopes which are now dominated by uncertainties in their respective $Q$-values. 
\end{abstract}

\maketitle

\section{I. Introduction}
Atomic nuclei composed of specific numbers of protons and neutrons known as magic numbers show particularly energetically favorable configurations. For example, the binding energy of these nuclei is significantly higher than what the semi-empirical mass formula~\cite{Benzaid2020}, derived from the liquid drop model of the nucleus, would predict. Understanding the microscopic origins of this phenomenon, known as nuclear shell evolution, is a central focus of contemporary nuclear physics \cite{SORLIN2008602,Kanungo_2013}.
The isotope $^{100}$Sn is often referred to as the `holy grail of nuclear structure research', as it is the heaviest known proton-bound nucleus among the self-conjugate ($N=Z$) nuclei, which is expected to be doubly magic with neutron and proton shell closures at $N=50$ and $Z=50$, respectively. Comprehensive reviews of the existing literature on $^{100}$Sn can be found in Refs.~\cite{FAESTERMANN201385,physics4010024}. Significant experimental efforts are underway to measure the properties of $^{100}$Sn. Its proximity to the proton drip-line makes it one of the rarest nuclei to study. As such, only its half-life, $\beta$-endpoint energy and mass are known so far~\cite{Hinke2012,PhysRevLett.77.2400}. Unfortunately, the limited precision of the latter does not provide sufficient clarity regarding the nuclear shell evolution along the proton drip-line. 

The properties of nuclei close to $^{100}$Sn are also of high interest for nuclear structure studies, see e.g. recent discussions in Refs.~\cite{Mougeot2021,PhysRevLett.131.022502,PhysRevC.102.014304,Reponen2021,HORNUNG2020135200,karthein2023}. The isotopes located northeast of $^{100}$Sn on the nuclear chart form an island of enhanced $\alpha$ and proton decay whose existence originates from the double magicity of $^{100}$Sn and the proximity to the proton drip-line~\cite{PhysRevC.85.014312}. They show a large Gamow-Teller $\beta$-decay strength~\cite{Hinke2012}, super-allowed $\alpha$ decay~\cite{PhysRevLett.121.182501, PhysRevLett.14.114, PhysRevC.73.061301,PhysRevLett.97.082501,Janas2005,PhysRevC.94.024314}, and potentially cluster~\cite{PhysRevC.52.726, PhysRevC.49.1922, PhysRevC.51.1762} and two-proton emission~\cite{PhysRevC.51.R1070, PhysRevLett.110.222501}. 
Moreover, these isotopes play a role in stellar nucleosynthesis as the astrophysical rapid proton (rp) capture process dies out in this region~\cite{PhysRevC.85.014312,AURANEN2019187,PhysRevLett.86.3471,PhysRevLett.102.252501,Grawe_2007,PhysRevLett.98.212501}. Precise knowledge of their binding energies, and thus their masses, are necessary to establish the energy balances of the relevant decays and transitions in this process. 

\begin{figure}[t]
\includegraphics[width=\columnwidth]{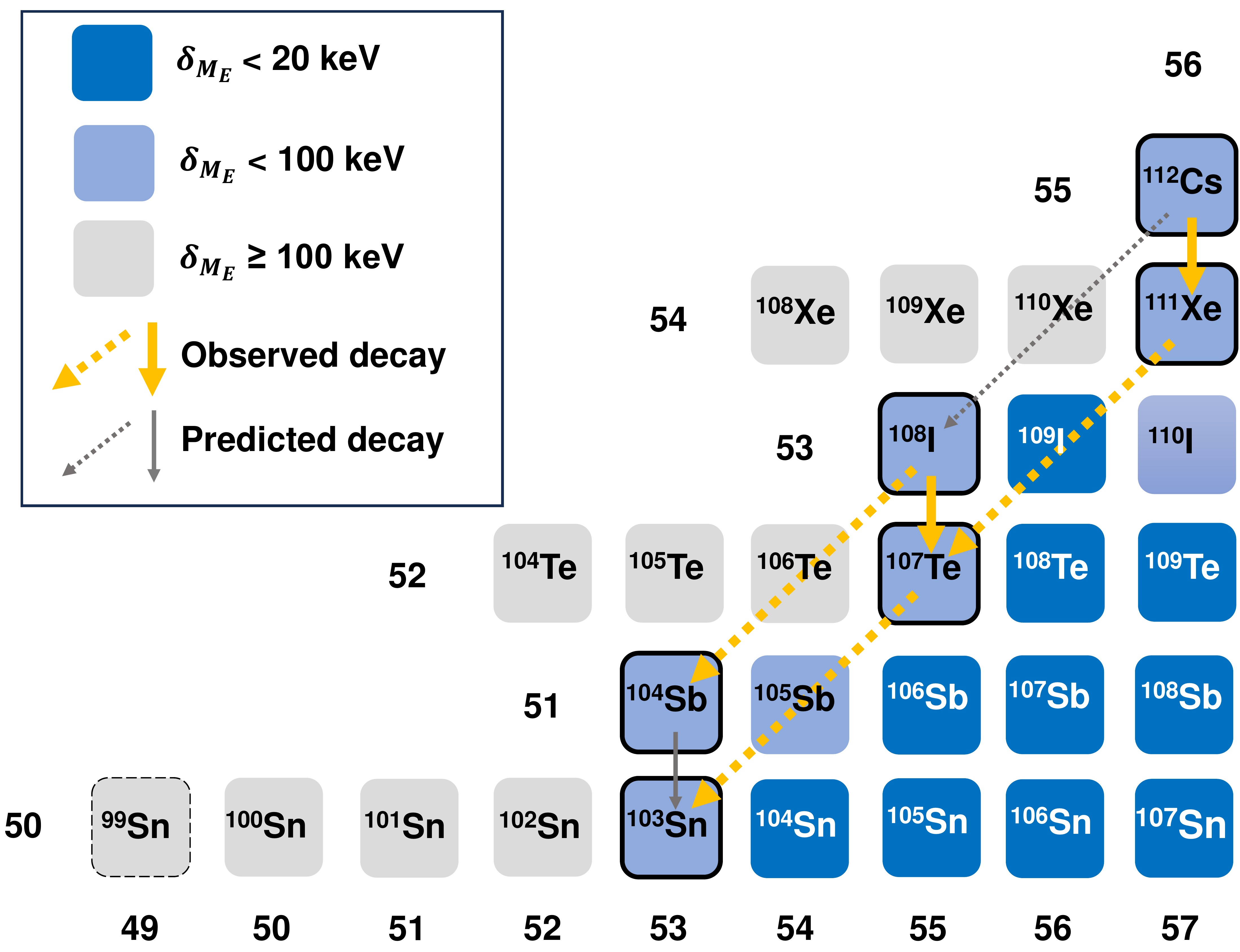}
\caption{The nuclear landscape in the region of $^{100}$Sn. Individual isotopes are colored based on their mass excess uncertainties $\delta_{M_E}$ prior to this work. A dashed border around an isotope refers to an extrapolated mass. The mass excesses of isotopes marked with a thick solid border were improved in this work. Directly observed and predicted decay chains are illustrated by thick orange and thin grey arrows, respectively, representing either proton (solid) or alpha (dashed) decay.} 
\label{fig:103Sn_Decay}
\end{figure}

Since the decay chains of $^{104}$Sb, $^{107}$Te, $^{108}$I, $^{111}$Xe, and $^{112}$Cs are terminated in $^{103}$Sn (see Fig.~\ref{fig:103Sn_Decay}), a precise knowledge of the mass of $^{103}$Sn is required to determine the masses of these five isotopes based on their respective $Q$-values known from decay spectroscopy~\cite{AME2020}. However, the mass of $^{103}$Sn has been a topic of debate.  The 2016 version of the Atomic Mass Evaluation (AME2016)~\cite{Wang_2017} reported an experimental mass excess of $-66970(70)$~keV for $^{103}$Sn derived from the $Q_{\beta^+}$ value and the mass of $^{103}$In, but significant inconsistencies with theoretical predictions, see e.g. Ref.~\cite{Mougeot2021}, and mass surface trends led to its replacement with an extrapolated value of $-67090(100)$~keV in AME2020~\cite{AME2020}. A subsequent direct mass measurement at the Cooler Storage Ring (CSRe) in Lanzhou in 2023 yielded a mass excess of $-67138(68)$~keV~\cite{StorageRing}, which is in good agreement with the extrapolated value in AME2020, restoring confidence in the smoothness of the mass surface. An independent and more precise mass measurement of $^{103}$Sn would further validate this finding and would serve as a crucial anchor for the masses of $^{104}$Sb, $^{107}$Te, $^{108}$I, $^{111}$Xe, and $^{112}$Cs. 

In this work, we report on high-precision Penning trap mass measurements of $^{103}$Sn performed at the Low Energy Beam and Ion Trap (LEBIT)~\cite{LEBIT} located at the Facility of Rare Isotope Beams (FRIB) using the time-of-flight ion cyclotron (ToF-ICR) technique~\cite{ToF1,BECKER199053,KONIG199595}. Beyond its relevance to nuclear structure and astrophysics, the high precision mass measurement of $^{103}$Sn serves as a step towards a measurement of $^{100}$Sn.  
With the ongoing ramp-up of beam intensity~\cite{doi:10.1142/S0217732322300063}, FRIB is expected to produce $^{100}$Sn, as well as many of its neighbors, at rates of a few ions per second in the near future, paving the way for exceptional research opportunities and deeper insights into the evolution of the nuclear shells in the vicinity of doubly magic $^{100}$Sn.

\section{II. Experimental Method and Analysis }
A beam of radioactive \textsuperscript{103}Sn was produced via projectile fragmentation. A $^{124}$Xe primary beam was accelerated in FRIB's superconducting Linear Accelerator \cite{York-LINAC} to an energy of 228 MeV/u and sent to a  2.015~mm thick $^{12}$C target, creating a cocktail of ion species that was passed to the Advanced Rare Isotope Separator (ARIS)~\cite{Hausmann-ARIS} for purification. In preparation for stopping the ions in a gas stopper, the momentum of the purified beam was compressed using an Al wedge of 1004~$\mu$m thickness at an angle of 2.67~mrad, followed by a 611~$\mu$m thick Al degrader. An optimal degrader angle of 17~degrees was determined adjusting the effective thickness seen by the incoming beam. This process prepared the beam for acceptance into the Advanced Cryogenic Gas Stopper (ACGS)~\cite{Lund-ACGS}, which utilizes helium buffer gas to stop the beam. After stopping, radio-frequency (RF) carpet surfing \cite{BOLLEN-IonSurfing} was used to guide the ions to an extraction orifice. After extraction, the ions passed through an RF quadrupole for differential pumping and were accelerated to 30~keV. A dipole magnet with resolving power $\approx$ 1500 selected all ion species with a mass-to-charge ratio $A/$Q=51.5, including \textsuperscript{103}Sn\textsuperscript{2+}. The choice of a half-integer $A/$Q was motivated by the increased purity of the beam delivered to LEBIT. At LEBIT, the continuous ion beam was injected into a linear buffer-gas-filled Paul trap cooler-buncher~\cite{CoolerBuncher} for accumulation, cooling and bunching of the ion beam. Following the extraction of the ions from the Paul trap as well-defined ion bunches with reduced emittance, the ions were guided into LEBIT's 9.4~T hyperbolic Penning trap~\cite{PenningTrap}. 

In the Penning trap, the ions were confined in three-dimensions by a homogeneous magnetic field~$B$ and a quadrupolar electrostatic field. The motion of an ion in a Penning trap is characterized by three eigenfrequencies: an axial frequency $\nu_{z}$ and two radial frequencies, $\nu_{-}$ (magnetron frequency) and $\nu_{+}$ (reduced-cyclotron frequency), where typically $\nu_{-}\ll\nu_{z}<\nu_{+}$. For an ideal trap, the cyclotron frequency $\nu_{c}$ of an ion can be approximated as the sum of its radial frequencies \cite{Gabrielse-Sideband},
\begin{equation}
\nu_{c} = \nu_{+} + \nu_{-}.
\end{equation}
Penning traps enable a precise determination of the mass~$m$ of the ion of interest by measuring its cyclotron frequency $\nu_{c} = qB/(2\pi m)$ relative to that of a well-known reference ion $\nu_{c,\mathrm{ref}} = q_{\mathrm{ref}}B/(2 \pi m_{\mathrm{ref}})$. The cyclotron frequency ratio $R$ is given by
\begin{equation}\label{eq:mass_from_ratio}
   R = \frac{ \nu_{c} }{ \nu_{c,\textrm{ref}} } = \frac{ q \cdot m_\textrm{ref} }{ q_\textrm{ref} \cdot m }  ,
\end{equation}
where $q$ and $q_{\mathrm{ref}}$ are the charges and $m$ and $m_{\mathrm{ref}}$ are the masses of the ion of interest and the reference ion, respectively. 

\begin{figure}[t]
\includegraphics[width=\columnwidth]{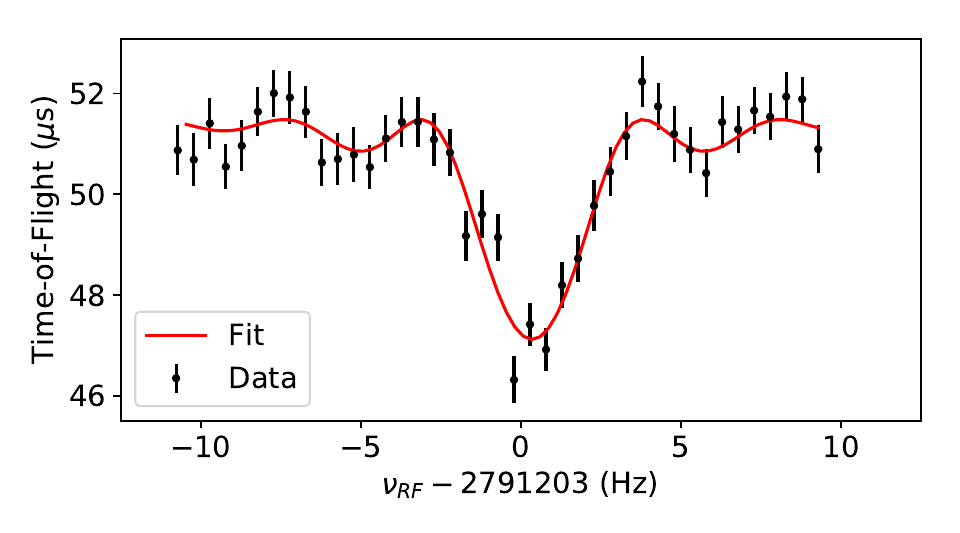}
\caption{A summed ToF-ICR spectrum of all 8 individual cyclotron frequency measurements taken for \textsuperscript{103}Sn\textsuperscript{2+} using the LEBIT 9.4 T Penning trap mass spectrometer. A $\chi^{2}$ minimization fit to the analytical curve described in \cite{KONIG199595}, depicted in red, was used to determine the frequency $\nu_{RF} = \nu_{c}$ occuring at the minimum time-of-flight. The width of the central dip corresponds to the inverse of the applied 250~ms quadrupolar excitation time.}
\label{fig:103Sn_ToFICR_summed}
\end{figure}

In this work, the mass of \textsuperscript{103}Sn\textsuperscript{2+} was determined using the ToF-ICR technique \cite{ToF1, KONIG199595,BECKER199053}. Extracted ions from the cooler-buncher were steered off-axis relative to trap center upon injection into the Penning trap using a Lorentz steerer \cite{LorentzSteerer}, inducing initial magnetron motion. The remaining isobaric contaminants at $A/$Q = 51.5 were cleaned using dipolar RF excitation applied to the central ring electrode near their respective reduced cyclotron frequencies \cite{DipoleCleaning}. A quadrupolar RF pulse applied for a chosen 250~ms of excitation time within a frequency range near that of the expected cyclotron frequency for \textsuperscript{103}Sn\textsuperscript{2+} converted the slow magnetron motion to fast reduced cyclotron motion. The ions were subsequently ejected and their time-of-flight to a microchannel plate (MCP) detector was recorded. Figure~\ref{fig:103Sn_ToFICR_summed} shows the ions' time-of-flight as a function of the applied quadrupolar frequency~$\nu_{RF}$. When $\nu_{RF}$ approached $\nu_{c}$, the ions' radial energy increased, resulting in a shorter flight time to the detector. This led to a time-of-flight minimum when $\nu_{RF}=\nu_{c}$. Measurements of the ion of interest for 250~ms of quadrupolar excitation time and measurements of the stable molecular reference $^{12}$\ce{C3}$^{1}$\ce{H2}$^{14}$\ce{N1}$^+$ for 500~ms of excitation time were interleaved to yield the average ratio $\bar{R}$, weighted by each measurement's uncertainty.  The mass of $^{103}$Sn was determined according to Eq.~\ref{eq:mass_from_ratio} accounting for the mass of the missing electron(s). The electron's binding energy itself is neglected, as it is smaller than the statistical uncertainty of $\bar{R}$.

\section{III. Results}
\begin{figure}[t]
\includegraphics[width=\columnwidth]{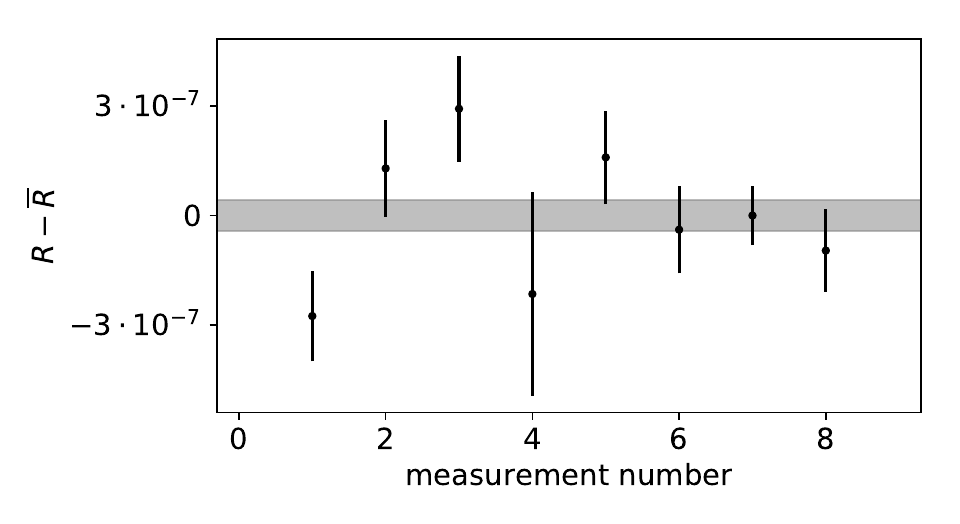}
\caption{Cyclotron frequency ratios $R$ with respect to the average ratio $\bar{R} =1.010779605(29)$. The gray bar shows the $\pm 1 \sigma$ uncertainty in $\bar{R}$.}
\label{fig:Ratios_103Sn}
\end{figure}

Eight ToF-ICR measurements of \textsuperscript{103}Sn\textsuperscript{2+} were taken over the course of approximately eight hours. The measured frequency ratios of the individual measurements are shown in Fig.~\ref{fig:Ratios_103Sn} relative to that of the stable reference. The weighted average mass ratio $\bar{R} = 1.010779605(29)$ corresponds to a mass excess of $-67125.9(3.7)$~keV for \textsuperscript{103}Sn. 

Several systematic effects add an uncertainty $\delta_{R}$ to~$\bar{R}$. Mass-dependent shifts related to inhomogeneity in the magnetic field and trap imperfections were studied at LEBIT in detail and are known to add an uncertainty of $\delta_{R} \approx 2 \times 10^{-10}/u$~\cite{MassOffsetError}.  
Additionally, non-linear temporal shifts in the magnetic field at LEBIT contribute $\delta_{R} < 10^{-9}$ per hour \cite{MagneticFieldShift}. To counter these, regular reference measurements were performed. The maximum ion cutoff $\leq 5$ also allowed the uncertainty due to ion-ion interactions to be $\delta_{R}\approx 10^{-8}$. 
These various uncertainties are negligible in comparison to the statistical precision of the measurement ($\approx 10^{-7}$).  The cyclotron frequency ratio~$R$ was periodically checked against the expected $R$ determined from masses obtained from AME2020 as measurements were performed to verify that it did not correspond to (molecular) isobars within the uncertainty of the measurement. No such isobars were plausible, providing confirmation that what was measured was indeed \textsuperscript{103}Sn\textsuperscript{2+}.

\section{IV. Discussion}
\begin{figure}[t]
\includegraphics[width=\columnwidth]{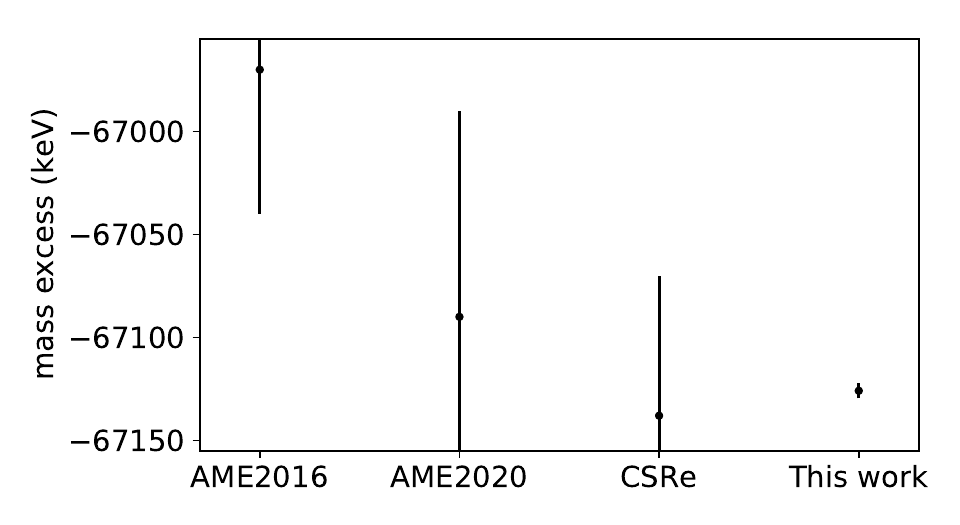}
\caption{Mass excess for $^{103}$Sn compared with literature (AME2016~\cite{Wang_2017}, AME2020~\cite{AME2020} and CSRe~\cite{StorageRing}).}
\label{fig:Comp_our_prev}
\end{figure}
\begin{figure}[t]
\includegraphics[width=\columnwidth]{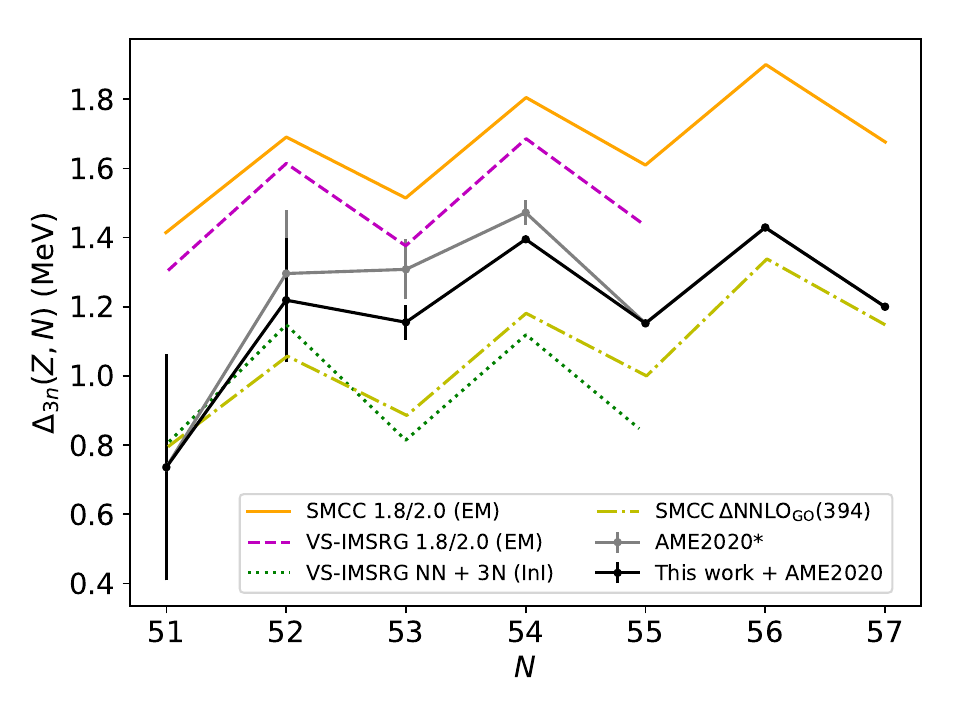}
\caption{Three-point estimator for the odd-even staggering for the tin isotopic chain ($Z=50$) as a function of neutron number $N$. The experimental data (grey and black curve) is taken from AME2020~\cite{AME2020} except for $^{103}$Sn. The black curve takes our new mass measurement of $^{103}$Sn into account. The grey curve utilizes the experimental mass of $^{103}$Sn as reported in AME2016~\cite{Wang_2017}, which was replaced with an extrapolated value in AME2020 . Beside of this mass of $^{103}$Sn the masses reported in AME2020 are used. The theoretical results of the valence-space formulation of the in-medium similarity renormalization group (VS-IMSRG) as well as shell-model coupled-cluster (SMCC) calculations are taken from Ref.~\cite{Mougeot2021}.}
\label{fig:Threepointestimator}
\end{figure}

Our mass value of $^{103}$Sn agrees with the previous measurement conducted at CSRe~\cite{StorageRing}, while improving the precision by more than an order of magnitude, see Fig.~\ref{fig:Comp_our_prev}. However, both of these measurements deviate from the mass reported experimentally in AME2016~\cite{Wang_2017}, which was derived indirectly from the known mass of $^{103}$In and the $\beta$-decay energy $Q_{\beta^+}$ between $^{103}$Sn and $^{103}$In~\cite{Wang_2017, MUKHA200466, Kavatsyuk2005}.  Assuming that the atomic mass $M$ of $^{103}$In is well known, we calculate a new $Q_{\beta^+}$ value based on our measured mass of $^{103}$Sn and the AME2020 mass of $^{103}$In~\cite{AME2020}, $Q_{\beta^+} = M (^{103}\mathrm{Sn})-M(^{103}\mathrm{In})$. Our updated $Q_{\beta^+}$ value is 7506(10)~keV, compared to the previous value of 7660(70)~keV. This discrepancy indicates a limited understanding of the full decay processes, possibly due to internal gamma conversion or undetected weak decay branches in Refs.~\cite{MUKHA200466, Kavatsyuk2005}. 
In AME2020~\cite{AME2020}, the mass of $^{103}$Sn was classified as a `seriously irregular mass' and replaced by an extrapolated value, which perfectly matches our new measurement. Consequently, the smoothness of the mass surface around $^{103}$Sn is confidently restored. 
Taking our new mass value of $^{103}$Sn into account, the experimental trend of the three-point estimator for the odd-even staggering $\Delta_{3n}(Z,N)= 0.5\cdot (-1)^N [M_E(Z, N-1) - 2M_E(Z,N) + M_E(Z,N+1)]$ along the tin isotopic chain is relatively well reproduced by various theoretical models that are discussed in detail in Ref.~\cite{Mougeot2021}, see Fig.~\ref{fig:Threepointestimator}. This increases confidence in the predictive power of these theoretical calculations for $^{100}$Sn.  

Furthermore, our new high-precision mass measurement of $^{103}$Sn significantly reduces the mass uncertainty of the alpha and proton emitters $^{104}$Sb, $^{107}$Te, $^{108}$I, $^{111}$Xe and $^{112}$Cs, which are located northeast of the nuclear chart with respect to $^{100}$Sn, see Fig.~\ref{fig:103Sn_Decay}. The mass excesses of these five isotopes were calculated based on our high-precision $^{103}$Sn measurement and the respective ~$Q$-values, see Tab.~\ref{tab: QTable}, 
\begin{equation}
\begin{split}
     M_E (^{107}\mathrm{Te}) &=  M_E (^{103}\mathrm{Sn}) + M_E (\alpha) + Q_{\alpha}(^{107}\mathrm{Te}), \\
     M_E (^{111}\mathrm{Xe}) &=  M_E (^{107}\mathrm{Te}) + M_E (\alpha) + Q_{\alpha}(^{111}\mathrm{Xe}), \\
     M_E (^{112}\mathrm{Cs}) &=  M_E (^{111}\mathrm{Xe}) + M_E (p) + Q_{p}(^{112}\mathrm{Cs}), \\
     M_E (^{108}\mathrm{I}) &=  M_E (^{107}\mathrm{Te}) + M_E (p) + Q_{p}(^{108}\mathrm{I}) \text{ and} \\
     M_E (^{104}\mathrm{Sb}) &=  M_E (^{108}\mathrm{I}) - M_E (\alpha) - Q_{\alpha}(^{108}\mathrm{I}). \\
\end{split}
\end{equation}
The respective mass excesses can be found in Tab.~\ref{tab: MassExcessTable}. 
Their uncertainties are now dominated by the uncertainties of the respective $Q$-values and not by the uncertainty in the mass excess of $^{103}$Sn. Since the astrophysical rp-process is terminated in this region~\cite{PhysRevC.85.014312,AURANEN2019187,PhysRevLett.86.3471,PhysRevLett.102.252501,Grawe_2007,PhysRevLett.98.212501}, these isotopes play a vital role in our understanding of stellar nucleosynthesis. Precise mass determinations, and ideally direct mass measurements in the future, aid in establishing the energy balances of the relevant decays and transitions.  

\begin{table}[H] 
\centering
\caption{Measured $Q$-values as reported in AME2020 \cite{AME2020} and used for the mass excess calculation of the five parent isotopes of $^{103}$Sn presented in Tab.~\ref{tab: MassExcessTable}.}
\vspace{\baselineskip}
\renewcommand{\arraystretch}{1.25}
\setlength{\tabcolsep}{12pt}
\begin{tabular}{l c c}
\hline
 & $Q$-value (keV) & references \\ \hline
$Q_{p}$($^{108}$I)    & 597(13) &  Ref.~\cite{AURANEN2019187}     \\ 
$Q_{p}$($^{112}$Cs)    & 816(4)   &     Refs.~\cite{PhysRevLett.72.1798,PhysRevC.85.034329}      \\ 
$Q_{\alpha}$($^{107}$Te)    & 4010(5)   &    Refs.~\cite{SCHARDT197965,Heine1991,AURANEN2019187,PhysRevC.101.014313}       \\ 
$Q_{\alpha}$($^{108}$I)     & 4099(5)    &    Refs.~\cite{PhysRevLett.72.1798,AURANEN2019187}     \\ 
$Q_{\alpha}$($^{111}$Xe)     & 3710(60)    &   Refs.~\cite{SCHARDT197965,SCHARDT1981153,Heine1991,PhysRevC.101.014313}       \\ 
\hline
\end{tabular}
\label{tab: QTable}
\end{table}

\begin{table}[H] 
\centering
\caption{Measured mass excess of $^{103}$Sn and calculated mass excesses based on the $^{103}$Sn measurement from this work (left) compared to the result from  CSRe~\cite{StorageRing} (right).}
\vspace{\baselineskip}
\renewcommand{\arraystretch}{1.25}
\setlength{\tabcolsep}{12pt}
\begin{tabular}{l c c}
\hline
 & $M_{E,\mathrm{LEBIT}}$ (keV) & $M_{E,\mathrm{CSRe}}$ (keV)  \\ \hline
$^{103}$Sn    & $-67125.9(3.7)$      & $-67138(68) $     \\ 
$^{104}$Sb    & $-59329(15)  $    & $-59340(70) $     \\ 
$^{107}$Te    &$-60691(6) $     & $-60700(70)  $    \\ 
$^{108}$I     & $-52805(14) $     &$ -52820(70)  $    \\ 
$^{111}$Xe     & $-54556(60)$      & $-54570(90) $     \\ 
$^{112}$Cs      &$ -46451(60)$     & $-46460(90) $     \\
\hline
\end{tabular}
\label{tab: MassExcessTable}
\end{table}

\section{V. Conclusions}
The first Penning trap mass measurement of $^{103}$Sn was performed at LEBIT~\cite{LEBIT} and a mass excess of $-67125.9(3.7)$~keV was obtained. This value improves the precision of the previous measurement~\cite{StorageRing} by more than an order of magnitude and restores confidence in the smoothness of the nuclear mass surface. Furthermore, the masses of five parent alpha/proton emitters calculated from this improved measurement of $^{103}$Sn have become well-anchored and their mass uncertainties are now dominated by their respective $Q$-values. The closed neutron and proton shells at $N=50$ and $Z=50$ of $^{100}$Sn provide a unique testing ground for our understanding of nuclear forces and shell evolution. Our measurement of $^{103}$Sn, just three neutrons away from $^{100}$Sn, lays the foundation for future high-precision mass measurements of $^{100}$Sn.

\section{Acknowledgements}
This material is based upon work supported by the U.S. Department of Energy, Office of Science, Office of Nuclear Physics and used resources of the Facility for Rare Isotope Beams (FRIB) Operations, which is a DOE Office of Science User Facility under Award Number DE-SC0023633.
This work was conducted with the support of Michigan State University, the US National Science Foundation under contracts nos. PHY-1565546 and PHY-2111185, the DOE, Office of Nuclear Physics under contract no. DE-AC02-06CH11357, DE-AC02-05CH11231, DE-SC0022538, and DE-SC0022538. 
S.E.C. acknowledges support from the DOE NNSA SSGF under DE-NA0003960.

\bibliography{Sn103.bib} 
\vspace{1mm}

\end{document}